\documentclass[doublecol]{epl2}

\title{Breakdown of the Kratky-Porod Wormlike Chain Model for
Semiflexible Polymers in Two Dimensions}
\author{Hsiao-Ping Hsu\inst{1} \and Wolfgang Paul \inst{2} \and K. Binder\inst{1}}

\shorttitle{Kratky-Porod Breakdown in Two Dimensions}
\institute{
  \inst{1}Institut f\"ur Physik, Johannes Gutenberg Universit\"at Mainz,\\
           Staudinger Weg 7, 55099 Mainz, Germany\\
  \inst{2}Theoretische Physik, Martin-Luther-Universit\"at Halle
  Wittenberg,\\ von Senckendorffplatz 1, 06120 Halle, Germany
}


\abstract{
By large-scale Monte Carlo simulations of semiflexible polymers in $d=2$ 
dimensions the applicability of the Kratky-Porod model is tested. This model 
is widely used as ``standard model'' for describing conformations and force 
versus extension curves of stiff polymers. It is shown that semiflexible 
polymers in $d=2$ show a crossover from hard rods to self-avoiding walks, 
the intermediate Gaussian regime (implied by the Kratky-Porod model) is 
completely absent. Hence the latter can also describe force versus extension 
curves only if the contour length is only a few times larger than the 
persistence length. Consequences for experiments on biopolymers at interfaces 
are briefly discussed.}

\begin{document}
\maketitle

Characterizing the flexibility or stiffness of polymer chains is of basic 
importance for describing their structure and dynamics, and hence 
relevant for understanding
the functions of biopolymers, as well as the application properties of 
synthetic polymers~\cite{1,2,3,4}. Moderately stiff (``semiflexible'') 
macromolecules behave like rods on small scales, and one captures this
behavior by the concept of the so-called ``persistence length'' $\ell_p$. For
larger length scales, entropic flexibility prevails and random coil-like
structures occur. Important examples for such stiff  
biopolymers are DNA, some proteins, actin, neurofilaments, but also mesoscopic 
objects such as viruses~\cite{5,6,7}. The experimental study of such 
biopolymers and the interpretation of these observations by models is a very 
active topic of research \{e.g.~\cite{8,9,10,11,12,13,14,15,16,17}\}. 
In particular, the conformation of these biopolymers can be directly visualized
by electron microscopy (EM) or scanning force microscopy (SFM) techniques when 
such polymers are adsorbed on substrates~\cite{8,9,10,12,14,17}; by atomic 
force microscopy (AFM) also force versus extension curves can be 
measured~\cite{11,13}. The same methods also work for synthetic polymers such 
as molecular brushes~\cite{18}, where stiffness is controlled by the length of 
side chains~\cite{19}.

The standard theoretical model, that is almost exclusively used 
\{e.g.~\cite{20,21,22,23,24,25,26,27,28,29,30,Toan}\} to interpret these experiments
is the simple ``wormlike chain (WLC) model''~\cite{31,32}. Its Hamiltonian is,
in the continuum limit,
\begin{equation} 
\label{eq1}
\frac{\mathcal{H}}{k_BT} = \frac{\kappa}{2} \int\limits^L_0 dt 
\Big(\frac{d^2 \vec{r} (t)}{d t ^2 } \Big)^2 \quad . 
\end{equation}
Here the curve $\vec{r}(t)$ describes the linear macromolecule, $t$ is a 
coordinate along its contour which has the length $L$. We choose units 
such that $k_BT=1$, and the bending stiffness $\kappa$ then is 
$\kappa=\ell_p/2$, in $d=2$ dimensions. In this paper we shall focus on 
the case of chains confined to two-dimensional geometry, since this case is 
relevant for the EM and SFM imaging techniques, and also the subject of 
numerous theoretical studies (e.g.~\cite{25,28,29,30}). However, the applicability
of eq.~(\ref{eq1}) in principle is questionable, since it neglects excluded 
volume between the repeat units of the chain completely. Thus, eq.~(\ref{eq1}) 
yields the end-to-end distance of the polymer chains as
\begin{equation} 
\label{eq2}
\langle R^2 \rangle = 2 \ell_p L \Big\{ 1- \frac{1}{n} [1-\exp(-n)] \Big\}
\quad , \quad n=L/\ell_p \quad,
\end{equation}
and hence for $n \gg 1$ the chain behaves like a Gaussian coil 
$( \langle R^2 \rangle= 2 \ell_pL)$ while for $n < 1$ it is essentially a 
rigid rod of length $L$. The bond-autocorrelation function shows then a simple 
exponential decay,
\begin{equation} 
\label{eq3}
g(t) = \langle \vec{a}_i \cdot \vec{a}_{i + s} \rangle = 
\ell_b^2 \exp (-t / \ell_p) \quad , \quad t= s \ell_b \quad,
\end{equation}
where we now consider a chain where $N_b$ bonds of length $\ell_b$ connect 
repeat units at sites $\vec{r}_i$, $\vec{a}_i=\vec{r}_{i + 1} - \vec{r}_i$, 
$| \vec{a}_i|= \ell_b$; so $L=N_b \ell_b$. Finally, if one considers the effect
of a force $f$ acting on one chain end (the other being fixed at the origin), 
by adding a term - $fX$ to the Hamiltonian ($X$ being the $x$-component of the 
end-to-end distance), one obtains from eq.~(\ref{eq1}) 
the force vs. distance
relation to a very good approximation, in $d=2$~\cite{29}

\begin{equation} \label{eq4}
f \ell_p =\frac{1}{8}[ 6 \frac{\langle X \rangle}{L} 
-1 + (1- \frac{\langle X \rangle }{L})^{-2} ] \quad .
\end{equation}
Since various experimental data have been described by 
eqs.~(\ref{eq2})-(\ref{eq4}) with some success adjusting parameters such as 
$\ell_p$ and $L$, it is widely believed that the basic Kratky-Porod model, 
eq.~(\ref{eq1}), describes semiflexible chains accurately, and a large body 
of work is concerned with various refinements of this model 
\{see e.g.~\cite{26,27,28,29,30}\}.

\begin{figure}
\begin{center}
\vspace{1cm}
(a)\includegraphics[scale=0.32,angle=270]{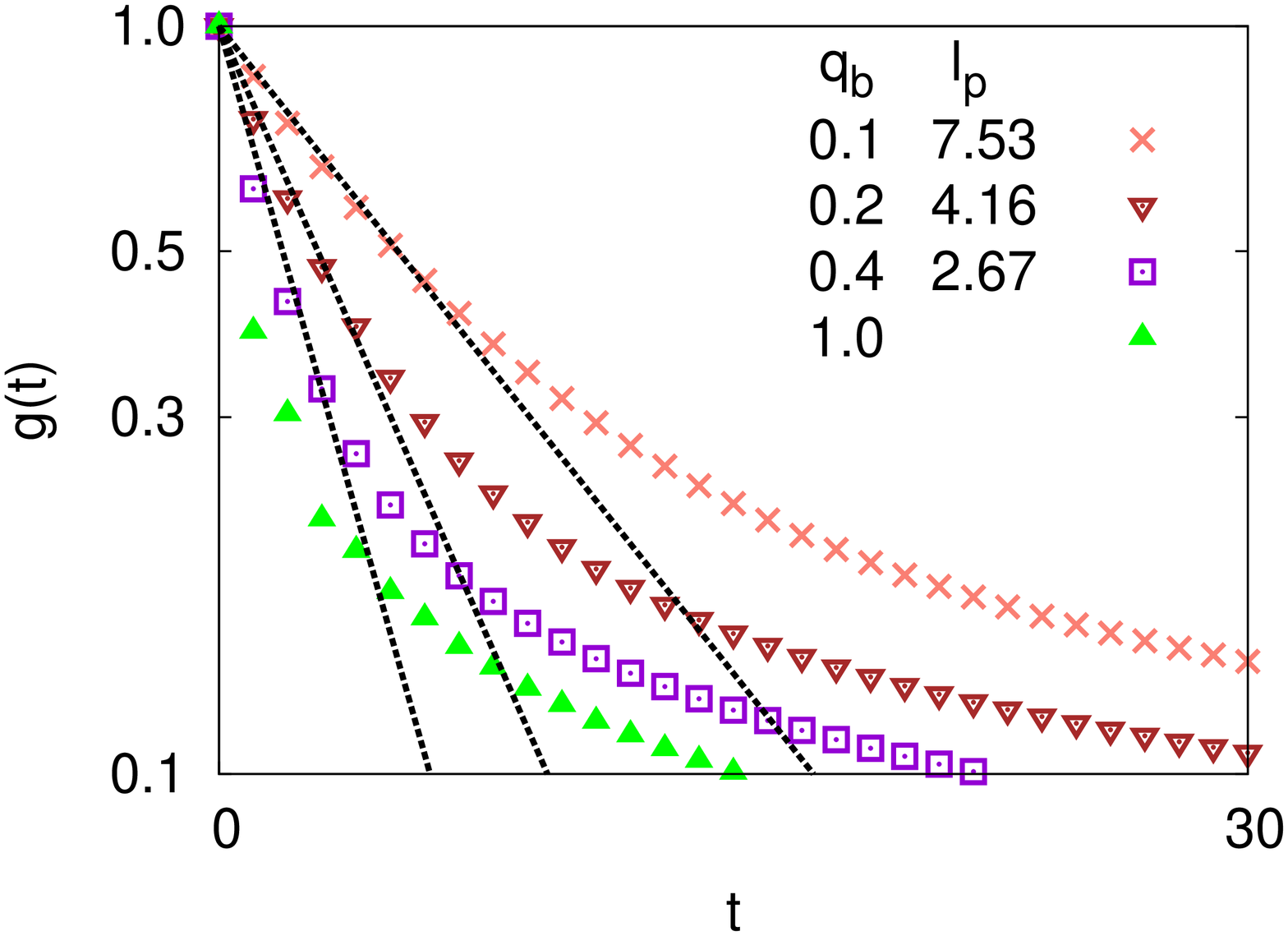}\\
(b)\includegraphics[scale=0.32,angle=270]{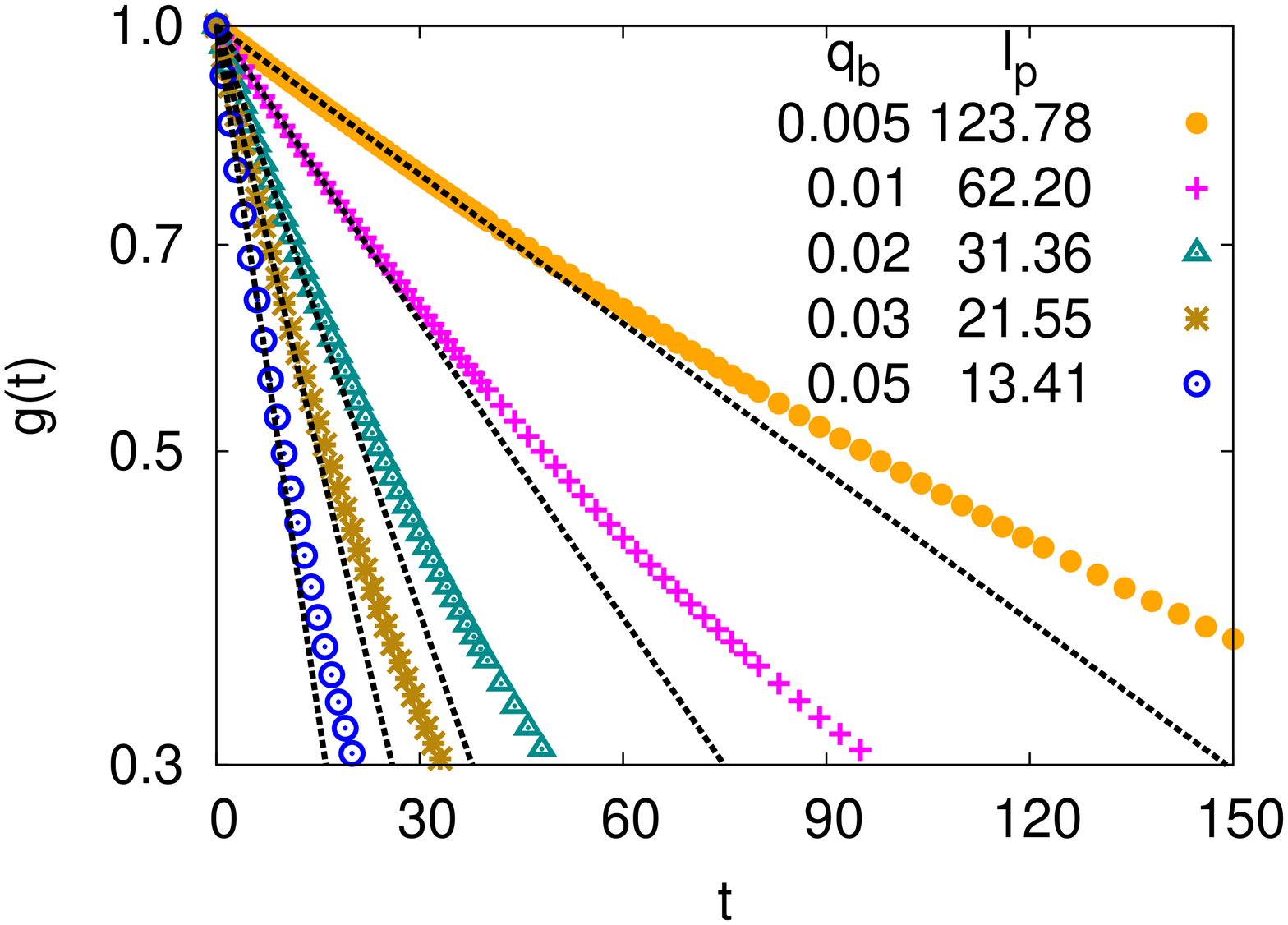}\\
(c)\includegraphics[scale=0.32,angle=270]{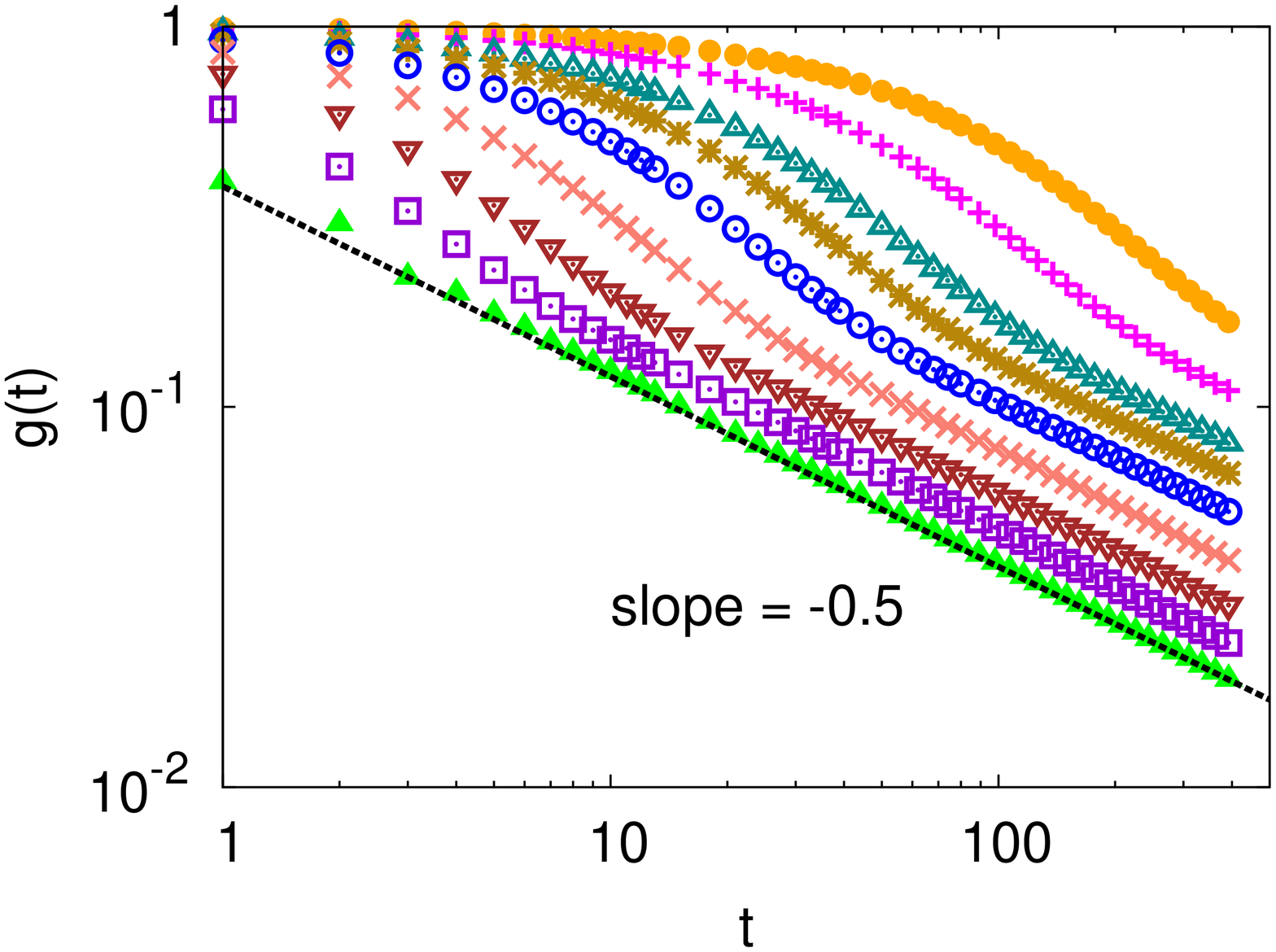}
\caption{Semi-log plots of the bond-correlation $g(t)$ vs. the contour length
$t$, for the ranges (for a definition of the parameters see eq.~(\ref{eq6})
  $0.1 \leq q_b \leq 1.0$ (a) and 
$0.005 \leq q_b \leq 0.05$ (b). The data are taken for $L=25600$ and $b=1$,
averaging over the site $i$ in eq.~(\ref{eq3}). Straight lines indicate fits
of the initial decay of $g(t)$ to eq.~(\ref{eq3}). The resulting values of
$\ell_p$ are quoted in the figure. (c) Log-log plot of $g(t)$ vs. $t$, for
$q_b=0.005$ to 1.0 (from above to below). The straight line shows a fit of
the data for $q_b=1$ and $t \geq 10$ to the power law
$g(t) \propto t^{-0.5}$.}
\label{fig1}
\end{center}
\end{figure}

However, in the present Letter we show that in fact in $d=2$ the validity of 
the Kratky-Porod model in the good solvent regime is very restricted, it 
always holds only up to contour lengths $L$ of a few times $\ell_p$, 
irrespective how large the persistence length $\ell_p$ is. In particular, 
a regime of $L$ where Gaussian statistics holds, 
$\langle R^2 \rangle = 2 \ell_pL$, in $d=2$ is completely absent, unlike the 
case of $d=3$, where for very large $\ell_p$ a double crossover 
(rods $\rightarrow$ Gaussian coils $\rightarrow$ non-Gaussian swollen coils) is established 
both 
experimentally~\cite{33} and theoretically~\cite{34}. Also eq.~(\ref{eq4}) 
breaks down for $L \gg \ell_p$, irrespective how large $\ell_p$ is. 
In $d=2$, we will show that
\begin{equation} 
\label{eq5}
\langle R^2 \rangle ^{1/2} \propto \ell^{1/4}_p L^{3/4} \quad, L > \ell_p
\end{equation}
and $g(t) \propto t^{-1/2}$, for $t > \ell_p$, rather than eq.~(\ref{eq3}). 
The latter result is consistent with the scaling prediction~\cite{35} 
$g(t) \propto t^{- \beta}$ with $\beta= 2 (1-\nu)$ where the Flory exponent 
$\nu=3/4$ in $d=2$, as written already in eq.~(\ref{eq5}).

There has been evidence for the scaling 
$\langle R^2 \rangle ^{1/2} \propto L^{3/4}$ for not so stiff polymers such 
as single stranded DNA in $d=2$ dimensions, see e.g.~\cite{9,10,16}, but it has
been widely believed that 
for very stiff polymers excluded volume interactions (that cause the 
nontrivial exponent $\nu=3/4$ rather than the Gaussian result $\nu=1/2$ which 
follows from eq.~(\ref{eq2})) can be
neglected, except for extremely long chains. We will show, however, that excluded 
volume effects set in strongly already for $L \approx 5 \ell_p$, invalidating the 
straightforward use of eqs.~(\ref{eq2})-(\ref{eq4}) for many cases of interest.

We carried out Monte Carlo simulations of self-avoiding walks (SAWs) on the 
square lattice, applying an energy $\varepsilon_ b$ if the orientation of bond 
vector $\vec{a}_i$ differs (by $\pm \pi/2$) from that of $\vec{a}_{i-1}$, and
using the pruned-enriched Rosenbluth method \cite{34,36,37}. The partition 
function of SAWs with $N_b$ steps and $N_{\rm bend}$ local bends is
\begin{equation} 
\label{eq6}
Z_{N,N_{\rm bend}} (q_b, b) =\sum\limits_{\rm config} C (N_b, N_{\rm bend}, X) 
q_b^{N_{\rm bend}}b^X \end{equation}
where $q_b=\exp (- \varepsilon_b/k_BT)$, $b=\exp (f /k_BT)$ and $X$ is the 
$x$-component of the end-to-end distance (assuming that the force $f$ acts in 
the $+x$-direction). 
In experiments where a force is applied to an end of a strongly adsorbed
chain, that takes essentially two-dimensional conformations, it is possible
to direct this force either perpendicular or parallel to the surface; only 
the latter case is considered here.
Note $q_b=1$ for flexible chains (standard SAWs) and 
$b=1$ in the absence of the force $f$. We generated data for 
$C(N_b, N_{\rm bend}, X)$ for $0.005 \leq q_b \leq 1.0$ and $N_b \leq 25600$.

Fig.~\ref{fig1} shows the bond-orientational correlations (for the case $f=0$).
For rather flexible chains, $q_b=0.4$, there are at best a few values 
$t=1$, $2$, $3$ compatible with an exponential decay (we use $\ell_b=1$ here 
and in the following). For small $q_b$, eq.~(\ref{eq3}) has a more extended 
range of applicability, and $\ell_p$ strongly increases when $q_b$ decreases, 
$\ell_p \approx 0.61/q_b$. But the asymptotic decay always is the expected 
power law (fig.~\ref{fig1}(c)). As has been emphasized 
recently~\cite{HsuMAC43}, in the presence of excluded volume ``the'' 
persistence length is a somewhat ill-defined concept; for the present
model, $\ell_p$ henceforth is defined from the initial slope of the curves
$\ln g(t)$ vs. $t$ as $t \rightarrow 0$.

\begin{figure}
\begin{center}
\vspace{1cm}
(a)\includegraphics[scale=0.32,angle=270]{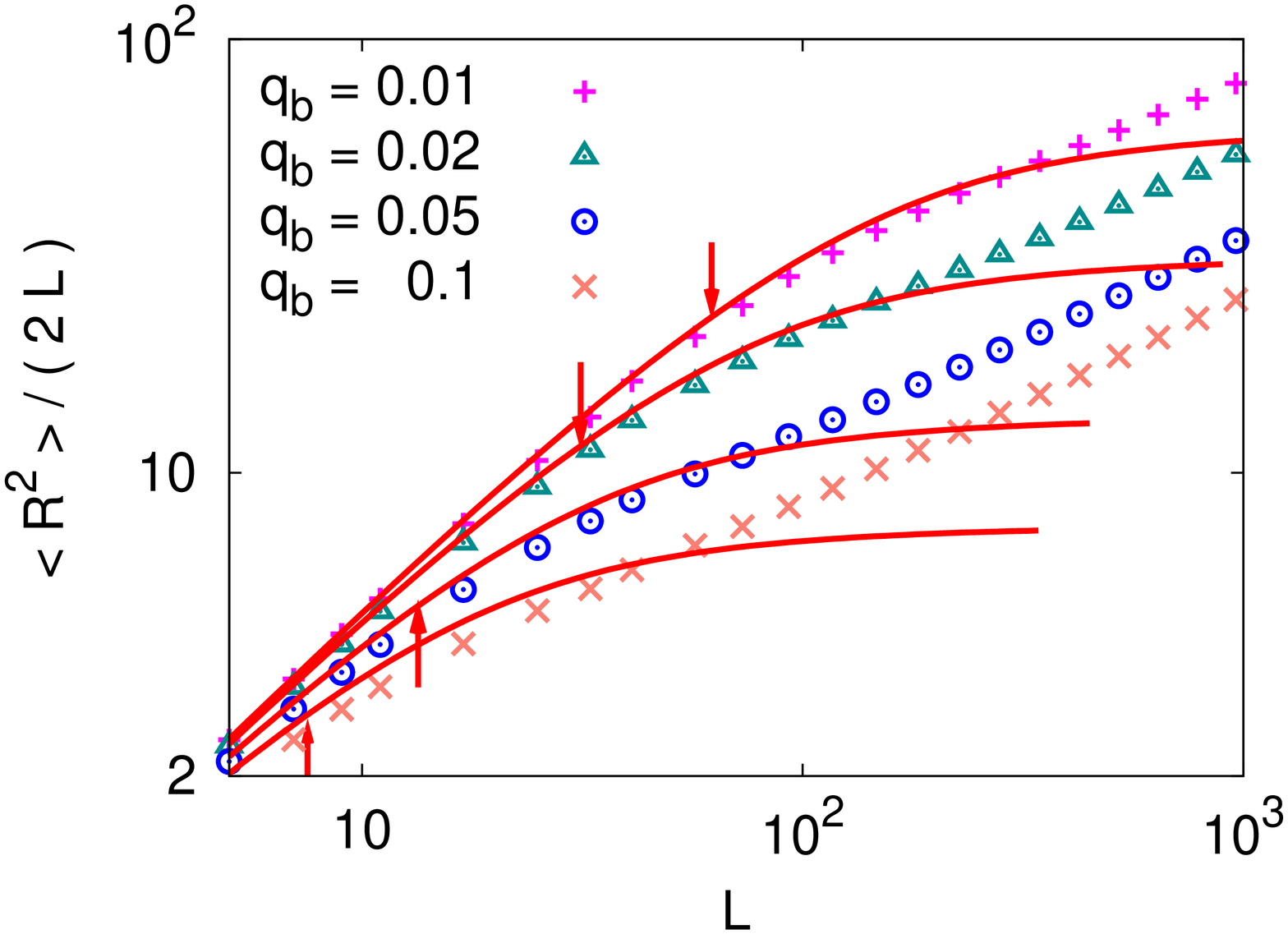} \\
(b)\includegraphics[scale=0.32,angle=270]{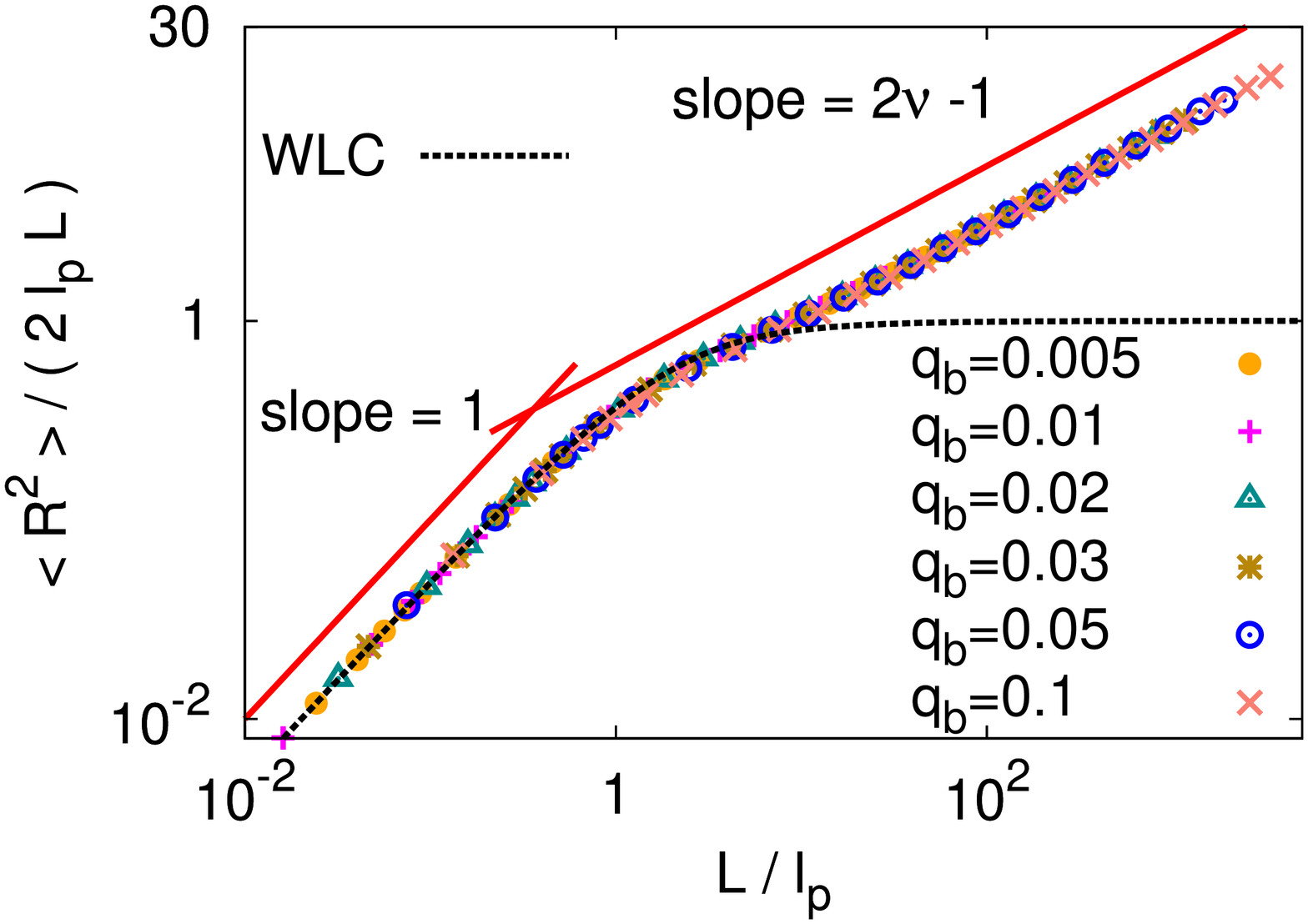}
\caption{Log-log plot of $\langle R^2 \rangle / (2L)$ versus
$L=N_b \ell_b$ (a) and log-log plot of $\langle R^2 \rangle / (2\ell_pL)$
versus $L/\ell_p$ (b), for $b=1$ and several choices of
$q_b$, as indicated. Full curves show the WLC prediction, eq.~(\ref{eq2}),
using $\ell_p$ (highlighted by arrows in (a)) from fig.~\ref{fig1}(a)(b) as an
input. Straight lines in (b) indicate the power laws in the rod regime
$(\langle R^2 \rangle =L^2$) and the SAW regime \{eq.~(\ref{eq5})\},
respectively.}
\label{fig2}
\end{center}
\end{figure}

\begin{figure}[htb]
\begin{center}
\vspace{1cm}
(a)\includegraphics[scale=0.32,angle=270]{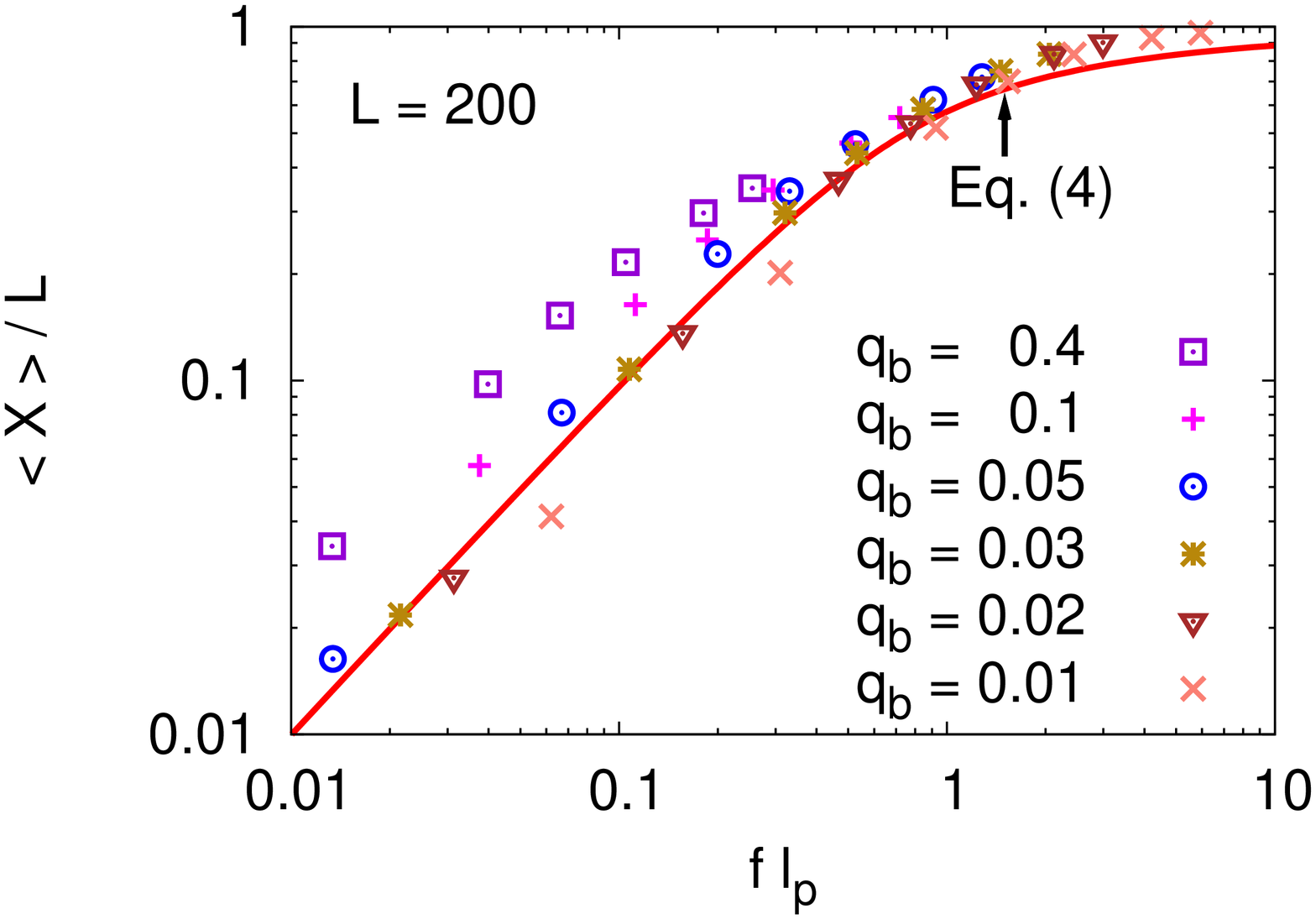} \\
(b)\includegraphics[scale=0.32,angle=270]{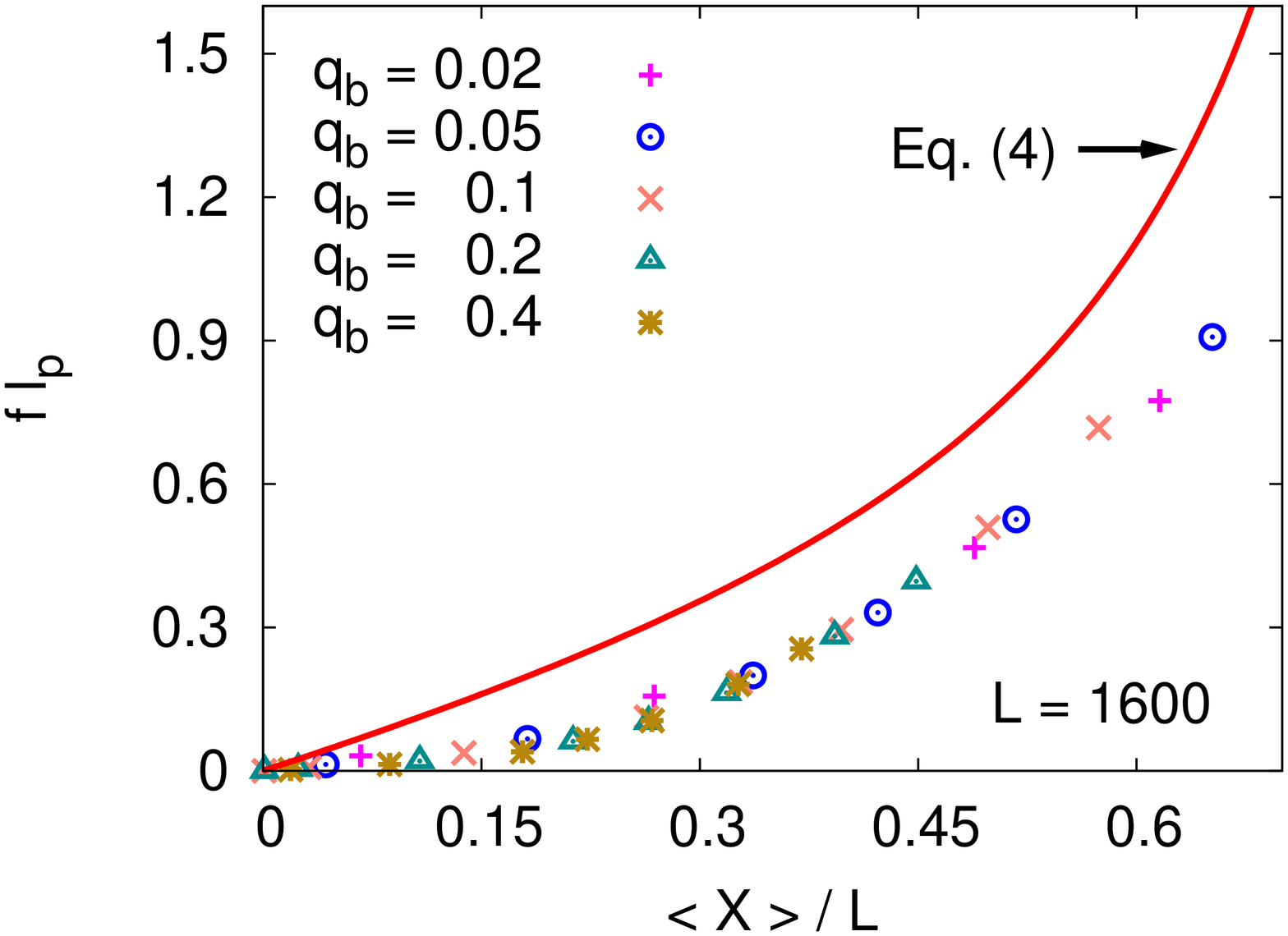} \\
(c)\includegraphics[scale=0.32,angle=270]{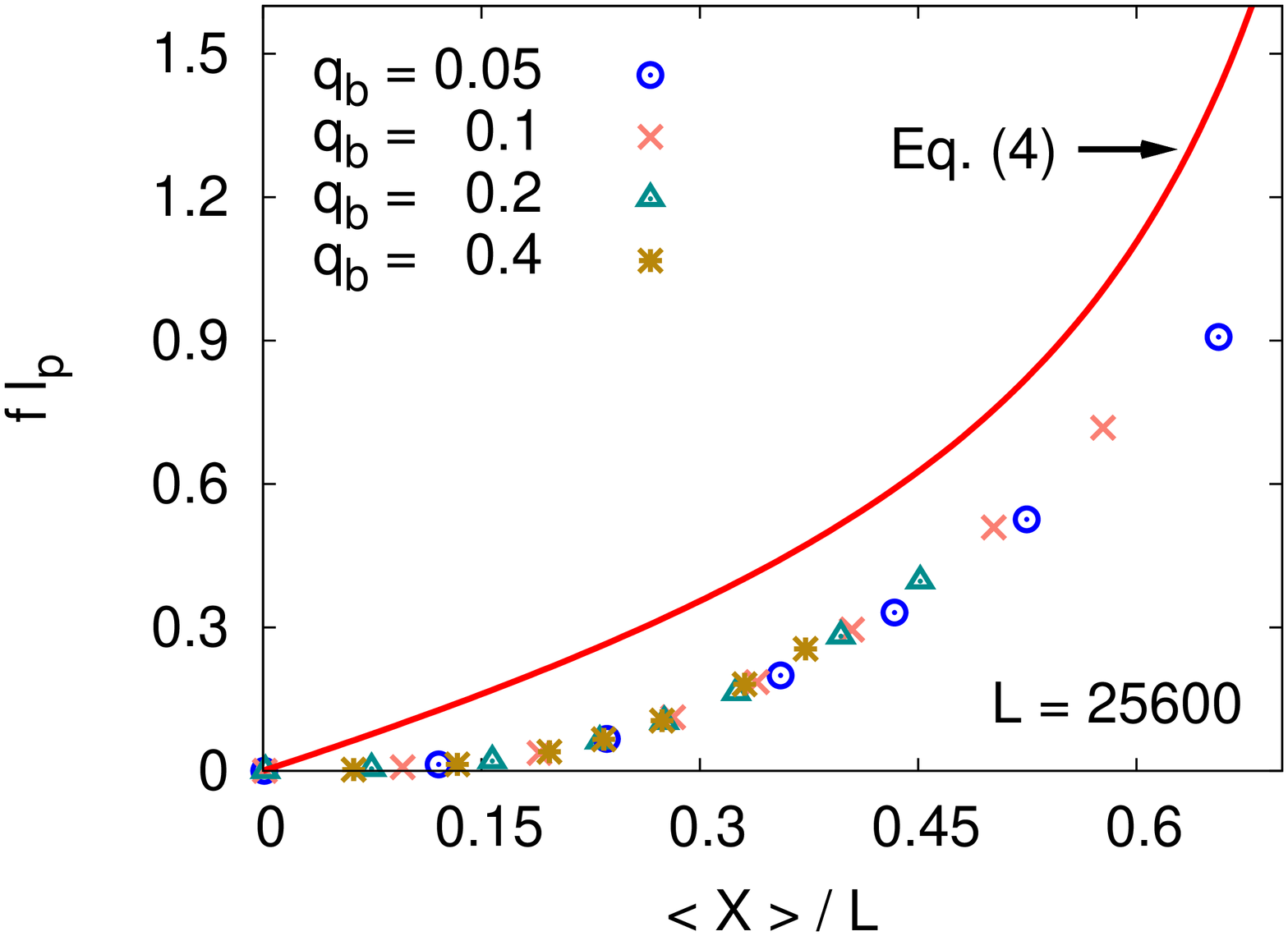}
\caption{Log-log plot of $\langle X \rangle /L$ 
vs. $f\ell_p$ for $L=200$.
Eq.~(\ref{eq4}) is shown by the full curve for comparison (a).
Rescaled force $f\ell_p$ plotted against $\langle X \rangle /L$
for $L=1600$ (b) and $L=25600$ (c). Various values of $q_b$ are shown
as indicated.}
\label{fig3}
\end{center}
\end{figure}

Fig.~\ref{fig2} presents a test of eq.~(\ref{eq2}). While eq.~(\ref{eq2}) 
trivially works for $L < \ell_p$ (the rod-like regime), significant deviations 
become visible for $L > 5 \ell_p$, irrespective of how large $\ell_p$ is, 
as the scaling plot (fig.~\ref{fig2}(b)) shows. In contrast to occasional claims 
in the literature~\cite{12}, a regime of Gaussian-like coils is completely 
absent in $d=2$. This result can be rationalized by the proper adaptation of 
Flory-type arguments~\cite{38} to $d=2$. The free energy of a stiff chain is 
taken as the sum of an elastic energy ($R^2 / \ell_pL)$ and the enthalpy due 
to repulsions, proportional to the 2nd virial coefficient 
$(\upsilon_2= \ell_p^2$~\cite{39}; prefactors of order unity are suppressed 
throughout)
\begin{equation} 
\label{eq7}
\Delta F = R^2 / (\ell_p L) + \upsilon_2 R^2 [(L/\ell_p)/R^2]^2 \quad .
\end{equation}
In $d=2$, the ``volume'' of a chain of radius $R$ scales like $R^2$, and 
the density of the $n=L/\ell_p$ subunits is $n/R^2$ in this volume. 
Minimizing $\Delta F$ with respect to $R$ yields eq.~(\ref{eq5}). The minimum
length $L$ where eq.~(\ref{eq5}) holds is found when the enthalpic term in 
eq.~(\ref{eq7}) is unity for $R^2= \ell_p L$, i.e. for 
$L^* = \ell_p^3/ \upsilon_2= \ell_p$, and there the rod-like regime starts: 
this argument shows that we should expect a single crossover from rods to SAWs,
as seen in fig.~\ref{fig2}(b), unlike the $d=3$ case \cite{34,38}.

How then can we understand the apparent success (suggested in the literature)
of the Porod-Kratky model to
analyze force-extension curves in 2d? In fig.~\ref{fig3} we show some of our
results on force vs. extension 
curves in $d=2$ and compare our data to the theoretical prediction based on 
the WLC model, eq.~(\ref{eq4}). Here the persistence length estimates
quoted in Figs.~\ref{fig1}(a)(b) were used, so we can compare our
simulation results that are based on eq.~(\ref{eq6}) to the prediction,
eq.~(\ref{eq4}), without adjusting any parameter whatsoever.
One can see that the latter equation works only approximately 
(fig.~\ref{fig3}(a)) for very stiff chains in a very restrictive range of 
contour lengths, where we can deduce from a detailed inspection of
the data that $6 < L/\ell_p <10$ must be fulfilled:
if $L/\ell_p$ is too small, the 
chain behaves as a flexible rod, which can be oriented by a force but not 
stretched; if $L/\ell_p$ is too large, excluded volume effects invalidate 
eq.~(\ref{eq4}), similarly as eq.~(\ref{eq2}) fails then. 
For $q_b=0.4$, the chains have hardly any rod-like regime as Fig.~\ref{fig1}(a)
reveals, $\ell_p$ is less than three lattice spacings, and so large deviation
from eq.~(\ref{eq4}) are no surprise, of course. For small $q_b$, where
for the chosen values of $L=200$ in Fig.~\ref{fig3}(a) $L$ is only a few
times larger than $\ell_p$ (recall $\ell_p \simeq 62$ for $q_b=0.01$,
Fig.~\ref{fig1}(b)), the deviations of the data from eq.~(\ref{eq4})
go into the opposite direction ($\langle X \rangle /L$ for $f\ell_p<1$
is smaller than predicted by eq.~(\ref{eq4}), while $\langle X \rangle /L$
is larger than predicted if $\ell_p$ is small). This finding implies that
for $L=200$ and intermediate values of $\ell_p$, the observed variation of
$\langle X \rangle /L$ with $f\ell_p$ is close to the predicted one,
for the intermediate range of $L/\ell_p$ quoted above, but this agreement is 
somewhat accidental.

Note also that a sensible test of the Kratky-Porod model (which is a 
continuum model) by our discrete lattice model is only possible for forces such
that $f\ell_p < 1$, since important deviations between discrete chain models and
the Kratky-Porod model occur~\cite{26} when the so-called deflection length
$\lambda \propto (f\ell_p)^{-1/2}$ of worm-like chains becomes smaller than 
the bond length $\ell_b$. Thus our data do not converge to eq.~(\ref{eq4}) even 
for large $f\ell_p$, although for very strongly stretched chains 
$(\langle X \rangle /L)$ close to unity) excluded volume effects must 
become irrelevant.

\begin{figure}[t]
\begin{center}
\vspace{1cm}
\includegraphics[scale=0.32,angle=270]{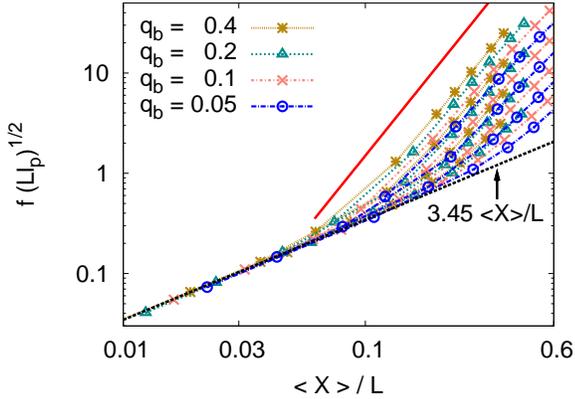}
\caption{Log-log plot of $f(L \ell_p)^{1/2}$ vs. $\langle X \rangle /L$,
including several values of $q_b$ as indicated, and data for
$N_b=400$, $1600$, $6400$, and $25600$ (from bottom to top at
the right side of the diagram, respectively). Straight dotted line (black) 
indicate the linear response behavior, and straight solid line (red) 
indicate the non-linear behavior, i.e. $f \propto \langle X \rangle ^3$ (see text).}
\label{fig4}
\end{center}
\end{figure}

   If $L$ is very large, such a crossing of the simulated curves for
$\langle X \rangle /L$ as function of $f\ell_p$ with eq.~(\ref{eq4})
when $\ell_p$ is varied does no longer occur (Fig.~\ref{fig3}(b)(c)).
The simulation results for $\langle X \rangle /L$ are now always significantly
larger than the prediction, eq.~(\ref{eq4}), particularly for small
values of $f\ell_p$. This huge discrepancy for small values of $f\ell_p$
can be understood readily in terms of a linear response argument:
actually, eq.~(\ref{eq4}) is found from adding a term $-fX$ to the Hamiltonian,
eq.~(\ref{eq1}). Therefore it is straightforward to derive, in the limit 
$f \rightarrow 0$, the linear response relation
\begin{equation}
\partial \langle X \rangle /\partial f \mid_{f=0} = \langle X^2 \rangle_{f=0} 
\,.
\label{eq8}
\end{equation}
Since $\langle X^2 \rangle_{f=0} = \langle R^2 \rangle /2$, where according to
the Kratky-Porod model \{eq.~(\ref{eq2})\} for $L \gg \ell_p$ we have simply
$\langle R^2 \rangle = 2\ell_pL$, we conclude that 
$\langle X \rangle = \langle X^2 \rangle f = \ell_p f L$
(in agreement with the Taylor expansion of eq.~(\ref{eq4}) to first order
in $\langle X \rangle /L$, as it must be, of course). However, in $d=2$
for vanishing force and $L \gg \ell_p$ the relation 
$\langle X^2 \rangle = \ell_p L$ must be replaced by 
$\langle X^2 \rangle \propto \ell_p^{1/2} L^{3/2}$, as is readily seen 
from eq.~(\ref{eq5}).Therefore we predict for the linear response regime
a very different scaling for the force-extension behavior, namely
\begin{equation}
   \langle X \rangle /L \propto \ell_p^{1/2} L^{1/2} f \, .
\label{eq9}
\end{equation}
This relation is tested in Fig.~\ref{fig4}. A wide range of choices of
contour lengths $L=N_b \ell_b$ and several choices of $q_b$ and hence
$\ell_p$ (the relation between $q_b$ and $\ell_p$ is quoted in 
Fig.~\ref{fig1}(a)(b)) are included. An interesting issue also is the regime
of relative extensions over which linear response holds:
while eq.~(\ref{eq4}) implies a linear
response regime applying almost up to $\langle X \rangle /L \approx 0.3$, 
irrespective of $\ell_p$, 
we suggest that the linear response breaks down if 
$\langle X \rangle^2 \approx \langle X ^2 \rangle$, i.e. for 
$\langle X \rangle /L \propto (\ell_p/L)^{1/4} \rightarrow 0$ as 
$\ell_p/L \rightarrow 0$. 
In the nonlinear regime, fig.~\ref{fig3}(b)(c) suggests that 
$\langle X \rangle /L$ can be 
described by some universal function of $\ell_pf$, that does not 
depend on $\ell_p$: 
this is the universality of $d=2$ SAWS, not the Kratky-Porod model.

Of course, the scaling $\langle X \rangle \propto L^{3/2}f$ for small $f$ is 
consistent with the scaling behavior proposed by Pincus~\cite{Pincus}
for stretched flexible polymers in the presence of excluded volume
\begin{equation}
  \langle X \rangle =R_0F(R_0/\xi_p)
\label{eq10}
\end{equation}
where $R_0$ is the radius of chain in the absence of a stretching force, 
$F(R_0/\xi_p)$ is a scaling function, and $\xi_p \propto 1/f$ is the size of 
``Pincus blobs'', and hence in the linear response regime 
$\langle X \rangle \propto R^2_0f$, i.e. eq.~(\ref{eq9}) results.
The condition that $\langle X \rangle / L$ is of order unity then leads 
to~\cite{Pincus} $\langle X \rangle \propto f^{1/\nu-1} = f^{1/3}$ in
$d=2$ dimensions, i.e, a strongly non-linear relation between $f$ and 
$\langle X \rangle$. This power law in fact is compatible with the data in 
Fig.~\ref{fig4} for large enough $\langle X \rangle / L$.

In conclusion, we have shown that in $d=2$ dimensions the 
Kratky-Porod model, that is
ubiquitously used to analyze the internal end-to-end distances of 
biopolymers such as DNA
\{e.g. \cite{15,17}\} or of synthetic polymers such as the bottle 
brushes \{e.g. \cite{18}\} and to analyze force versus extension 
curves \{e.g. \cite{11,13}\} has a very limited validity: it trivially 
describes the rod-like regime, $L \leq \ell_p$, but this regime is not 
useful in the context of such measurements, which are devoted to 
understanding the dependence of the persistence length on various 
parameters (such as particular amino acid sequences in DNA, or side 
chain length in bottle brushes, etc.). In $d=2$, a regime where 
Gaussian statistics (requiring $L\gg \ell_p$) holds is 
completely absent.     

Our findings imply that conformations of semiflexible polymers
in $d=2$ (equilibrated surface adsorbed case) depend on their relative
length $L/\ell_p$ very differently from the case $d=3$ (dilute
bulk solution). Thus there is no direct way to infer properties (such as 
$\ell_p$) in the bulk from measurements on surface adsorbed chains:
there is no simple relation between the effective persistence lengths
either \{in our model $\ell_p \propto 1/(4q_b)$ for $d=3$
but $\ell_p \cong 0.61/q_b$ in $d=2$ for $q_b \rightarrow 0$ \}. 

Going beyond the strictly 2-dimensional case, 
exploring the crossover to weak adsorption 
(chains with dangling non-adsorbed ``tails"
and ``loops" in addition to adsorbed ``trains") will be intriguing.
Also, the effects of excluded volume on force versus extension
curves when strongly adsorbed chains are pulled off a surface
in the direction normal to the surface by an AFM tip need to be
studied carefully. Thus, much further work is needed for a better
modeling of biopolymers and other stiff polymers at interfaces,
and on the interpretation of the corresponding experiments.

The effects studied in our work should also be relevant when one
studies semiflexible chains confined to the surface of a sphere
or its interior~\cite{Morrison}, a problem believed to be of great 
biological relevance.

\underline{Acknowledgements}: We thank the Deutsche Forschungsgemeischaft(DFG)
for support (grant No SFB625/A3) and the J\"ulich Supercomputing
Centre (JSC) for computer time at the NIC Juropa supercomputer.

\end{document}